# An investigation of factors affecting high-precision Sr isotope analyses ($^{87}Sr/^{86}Sr$ and $\delta^{88/86}Sr$) by MC-ICP-MS

Jiuyuan Wang[*], Dan Asael, Noah J. Planavsky, Lidya G. Tarhan

Department of Earth and Planetary Sciences, Yale University, 210 Whitney Ave, New Haven, CT 06511, USA

*Corresponding author: Jiuyuan Wang

Email address: jiuyuan.wang@yale.edu

# An investigation of factors affecting high-precision Sr isotope analyses ($^{87}$Sr/$^{86}$Sr and $\delta^{88/86}$Sr) by MC-ICP-MS

**Jiuyuan Wang*, Dan Asael, Noah J. Planavsky, Lidya G. Tarhan**

Department of Earth and Planetary Sciences, Yale University, 210 Whitney Ave, New Haven, CT 06511, USA

*Corresponding author: Jiuyuan Wang

Email address: jiuyuan.wang@yale.edu

**ABSTRACT**

The abundance and fractionation of the stable strontium (Sr) isotope system are being increasingly utilized to move forward our understanding in geological and cosmological processes. Two analytical techniques are commonly used to measure stable Sr isotopes: 1) double-spike thermal ionization mass spectrometry (DS-TIMS) and 2) Zr-doped sample-standard bracketing multi-collector inductively coupled plasma mass spectrometry (Zr-doped SSB via MC-ICP-MS). Relative to DS-TIMS, Zr-doped SSB via MC-ICP-MS allows simultaneous determinations of both $^{87}$Sr/$^{86}$Sr and $^{88}$Sr/$^{86}$Sr ratios, increasing measurement efficiency and sample throughput. However, this technique is currently associated with greater uncertainties in measurement precision and accuracy. In this study, we evaluated potential factors that can affect the quality of Sr isotope measurements during Zr-doped SSB. Our tests show that incomplete Sr recovery during chromatographic separation, mismatches of Sr and Zr concentrations and acid molarity between sample and bracketing standard, and cation contamination could all affect the precision and accuracy of Sr isotope measurements. We present evidence that, with updated preparation procedures and diligent concentration checks, a long-term reproducibility (2$\sigma_{SD}$: $^{87}$Sr/$^{86}$Sr =

±0.000015 and δ$^{88/86}$Sr = ±0.03 ‰) comparable to that of DS-TIMS is achievable when using the Zr-doped SSB method via MC-ICP-MC.

## 1. INTRODUCTION

Strontium (Sr) is one of the most extensively studied isotope systems in Earth and planetary sciences. Of its four stable isotopes ($^{84}$Sr, $^{86}$Sr, $^{87}$Sr, and $^{88}$Sr), $^{87}$Sr has two sources—a primordial origin that formed during nucleosynthesis and radiogenic input generated by radioactive $^{87}$Rb β-decay. Analysis of Sr isotopes has been carried out for several decades (e.g., Papanasyassiou and Wasserburg, 1969; Faure and Powell, 1972). Yet, these measurements have mainly been focused on the abundance of radiogenic $^{87}$Sr, which is reported as the $^{87}$Sr/$^{86}$Sr ratio. Due to limitations of measurement precision, the normalization procedure typically employed to account for instrumental mass discrimination eliminates any naturally occurring mass-dependent fractionation (e.g., Faure and Mensing, 2005). Thus, the natural fractionation of Sr isotopes has been largely overlooked, despite the extensive applications of $^{87}$Sr/$^{86}$Sr ratio.

Advancements in mass spectrometry have enabled the measurement of stable Sr isotope ratios, $^{88}$Sr/$^{86}$Sr, shedding new light on the mass-dependent fractionation of Sr isotopes during cosmological and geological processes (e.g., Patchett, 1980; Fietzke and Eisenhauer, 2006; Ohno and Hirata, 2007). The $^{88}$Sr/$^{86}$Sr ratio, which is reported in δ-notation

$$\delta^{88/86}\text{Sr} = \left[\frac{\left(\frac{^{88}\text{Sr}}{^{86}\text{Sr}}\right)_{\text{Sample}}}{\left(\frac{^{88}\text{Sr}}{^{86}\text{Sr}}\right)_{\text{NIST SRM 987}}} - 1\right] \times 1000,$$

has been recognized as a powerful proxy to gain insights into C and Sr cycles (Halicz et al., 2008; Krabbenhöft et al., 2009; Moynier et al., 2010; Charlier et al., 2012; Stevenson et al., 2014; Vollstaedt et al., 2014; Voigt et al., 2015; Pearce et al., 2015; Andrews et al., 2016; AlKhaitb and Eisenhauer, 2017; Wang et al., 2019; Shao et al., 2020; Klaver et al., 2020; Wang et al., 2021; Paytan et al., 2021). In general, two methodologies have been developed for the analysis of $\delta^{88/86}$Sr: multi-collector thermal ionization mass spectrometry using double-spike technique (DS-TIMS; Patchett, 1980; Krabbenhöft et al., 2009; Neymark et al., 2014; Andrews et al., 2016) and multi-collector inductively coupled plasma mass spectrometry (MC-ICP-MS) using Zr-doped sample-standard bracketing technique (Zr-doped SSB; Ohno and Hirata, 2007; Yang et al., 2008; Liu et al., 2012; Ma et al., 2013). The DS-TIMS method can achieve high precision and accuracy (e.g., $2\sigma_{SD}$: $^{87}$Sr/$^{86}$Sr = ±0.00001 and $\delta^{88/86}$Sr = ±0.02 ‰), but it requires separate purification and analytical procedures for both $^{87}$Sr/$^{86}$Sr and $\delta^{88/86}$Sr. On the other hand, the Zr-doped SSB method retrieves both $^{87}$Sr/$^{86}$Sr and $\delta^{88/86}$Sr from a single measurement, permitting higher sample throughput, but the precision is usually lower than the DS-TIMS method (e.g., $2\sigma_{SD}$: $^{87}$Sr/$^{86}$Sr = ±0.00004 and $\delta^{88/86}$Sr = ±0.04 ‰). This difference in precision can potentially result from a number of factors, including fractionations in chromatographic columns, isobaric interferences, and matrix effect.

Here, we conducted a systematic study to evaluate potential factors affecting the accuracy and precision of Sr isotope analyses using the Zr-doped SSB method on MC-ICP-MS. Our investigation shows that isobaric interreference, incomplete recovery from the Sr purification process, mismatches of Sr and Zr concentrations and acid molarity between sample and standard, and contamination in the solution matrix can all lead to large analytical artifacts and therefore need to be systematically monitored.

## 2. METHODOLOGY

### 2.1 Sample preparation, purification, and Zr doping

Four international standards (NIST SRM 987, IAPSO Standard Seawater, USGS BHVO-2, and USGS BCR-2) were used in the present study. The NIST SRM 987 standard (SrCO$_3$) was dissolved in 5% HNO$_3$ and used as the primary reference standard and the bracketing standard in the SSB. The digestion protocols for basalt and carbonate standards followed the procedures described in Andrews et al. (2016) and Wang et al. (2019), respectively. Briefly, for each basalt standard, approximately 100mg of powder were reacted by ~20 mL 5% $HNO_3$ in an acid-cleaned centrifuge tube and sonicated for 10 min to remove the potential carbonate fraction. After centrifuging, the supernatant was discarded. The remaining residue was washed with Milli-Q water into a Teflon beaker, dried, and then reacted in a 5:3 mixture of concentrated HF and $HNO_3$ at 130°C for 3 days. The mixture was then gently dried down at 80°C, refluxed in 4:1 mixture of concentrated HCl and $HNO_3$ at 80°C for 12 hours, dried, and finally re-dissolved in 5% $HNO_3$. For each carbonate standard, ~100 mg of powder was dissolved in ~20 mL of 5% $HNO_3$ in an acid-cleaned centrifuge tube and gently shaken on a rocker table overnight to ensure complete reaction. The mixture was centrifuged, and the supernatants were passed through an acid-cleaned 0.45 μm polypropylene syringe filter, collected in a Telfon beaker, dried at 80°C, and re-dissolved in 5% $HNO_3$ for column chemistry.

For the chromatographic purification of Sr, aliquots of dissolved samples containing 500 ng of Sr were weighted into Teflon vials, dried, and re-dissolved in 1 mL 8N $HNO_3$. Bio-Rad columns (Cat. No. 732-6008) were packed with ~1 mL Eichrom Sr-Spec resin (100-150 μm), cleaned by alternating between 8N $HNO_3$ and Milli-Q water six times, and then re-equilibrated with 8N $HNO_3$ before loading samples. To ensure separation performance and Sr yield (see discussion in Section 3.2), eight reservoirs (~8 mL) of 8N $HNO_3$ and four reservoirs (~4 mL) of 3N $HNO_3$ were eluted through each column. The purified Sr were collected with six reservoirs (~6 mL) of Milli-Q water into a beaker. The Sr solutions were dried, reacted with $H_2O_2$ to remove any potential organics, refluxed with concentrated $HNO_3$, re-dried, and finally re-dissolved with 5% $HNO_3$ for isotopic analyses. Total procedural blanks during the period of this study were ~150 pg for Sr (n=8), which is negligible compared to the amounts of sample Sr processed for isotopic analyses.

The Zr solution for doping was prepared from a Zr standard solution (SCP Science PlasmaCAL Zr calibration standard—1000 ppm in 5% $HNO_3$). 0.1 mL of the Zr standard solution were dried in a Teflon beaker, re-dissolved in 5% $HNO_3$, and diluted to a solution with Zr concentration of 400 ng/g. Besides samples for Sr and Zr concentration matching, the bracketing standard NIST 987 and all other samples were prepared in 5% $HNO_3$ solutions with concentrations of 100 ng/g for Sr and 200 ng/g for Zr.

### 2.2 Mass spectrometry and Sr isotope ratios determination

Radiogenic and stable Sr isotope ratios ($^{87}Sr/^{86}Sr$ and $^{88}Sr/^{86}Sr$) were measured on a Thermo Neptune Plus MC-ICP-MS equipped with $10^{12}$ Ω resistors at the Yale Metal Geochemistry Center. The samples were introduced through an ESI Apex Omega high-sensitivity sample introduction system, which consists of a heated cyclonic spray chamber, multi-stage Peltier-cooled condenser, and a helical EPTFE fluoropolymer membrane desolvator. Nitrogen was used as an additional gas to increase signal stability. Signals of nine masses (83, 84, 85, 86, 87, 88, 90, 91, and 92) were passed through Nickel cones and collected simultaneously for Sr and Zr isotope ratios as well as Kr and Rb isobaric interferences. The detailed cup configuration is listed in Table 1. Each sample and the bracketing standard were measured for 50 cycles with an integration time of 4.194 s in each cycle for both Sr and Zr. For each analytical session, electrostatic lens settings were adjusted to maximize the signal intensity on mass 88. Instrumentation details and operating parameters are also summarized in Table 1.

Rinse solution (5% $HNO_3$) was repeatedly measured in between each sample and bracketing standard. The values of nine measured masses were subtracted by the average of two adjacent rinse solutions to remove any potential background noise and contamination. After background correction, two additional corrections were made for $^{86}$Kr and $^{87}$Rb isobaric interferences. The $^{86}$Kr was determined by using the measured $^{83}$Kr intensity and a fixed conventional $\frac{^{86}Kr}{^{83}Kr}$ value:

$$^{86}Kr_{calculated} = {}^{85}Kr_{measured} \times \left(\frac{^{86}Kr}{^{83}Kr}\right)_{ref} \quad (1)$$

where "calculated" and "measured" denote the calculated $^{86}$Kr and measured $^{85}$Kr intensities. The $\left(\frac{^{86}Kr}{^{83}Kr}\right)_{ref}$ is the international convention value of 1.493955 (IUPAC, De Laeter et al., 2003). Following subtraction of calculated $^{86}$Kr, background-corrected $^{86}$Sr values were used to calculate $\frac{^{88}Sr}{^{86}Sr}$ ratios. The $^{87}$Rb correction was performed using the measured $^{85}$Rb, based on the exponential law:

$$^{87}Rb_{calculated} = {}^{85}Rb_{measured} \times \left(\frac{^{87}Rb}{^{85}Rb}\right)_{ref} \times \left[\left(\frac{^{88}Sr}{^{86}Sr}\right)_{ref} \Big/ \left(\frac{^{88}Sr}{^{86}Sr}\right)_{sample}\right]^{\ln\left(\frac{m87}{m85}\right)/\ln\left(\frac{m88}{m86}\right)} \quad (2)$$

Where "m" is the mass of responding isotopes, $\left(\frac{^{88}Sr}{^{86}Sr}\right)_{sample}$ is calculated using background-corrected $^{88}$Sr and Kr-corrected $^{86}$Sr, and $\left(\frac{^{87}Rb}{^{85}Rb}\right)_{ref}$ and $\left(\frac{^{88}Sr}{^{86}Sr}\right)_{ref}$ are the international convention values of 0.385697 and 8.375209, respectively (IUPAC, De Laeter et al., 2003). The $^{87}$Rb$_{calculated}$ were then subtracted from the background-corrected $^{87}$Sr values to remove $^{87}$Rb isobaric interferences.

The instrumental mass discrimination of Sr isotopic ratio measurements was corrected by Zr using the exponential law. The corrected Sr isotope ratios, $\left(\frac{^{87}Sr}{^{86}Sr}\right)_{corr}$ and $\left(\frac{^{88}Sr}{^{86}Sr}\right)_{corr}$, were determined using the following equations:

$$\left(\frac{^{87}Sr}{^{86}Sr}\right)_{corr} = \left(\frac{^{87}Sr}{^{86}Sr}\right)_{sample} \times \left[\left(\frac{^{92}Zr}{^{90}Zr}\right)_{ref} \Big/ \left(\frac{^{92}Zr}{^{90}Zr}\right)_{measured}\right]^{\ln\left(\frac{m87}{m86}\right)/\ln\left(\frac{m92}{m90}\right)} \quad (3)$$

$$\left(\frac{^{88}Sr}{^{86}Sr}\right)_{corr} = \left(\frac{^{88}Sr}{^{86}Sr}\right)_{sample} \times \left[\left(\frac{^{92}Zr}{^{90}Zr}\right)_{ref} \Big/ \left(\frac{^{92}Zr}{^{90}Zr}\right)_{measured}\right]^{\ln\left(\frac{m88}{m86}\right)/\ln\left(\frac{m92}{m90}\right)} \quad (4)$$

where $\left(\frac{^{87}Sr}{^{86}Sr}\right)_{sample}$ and $\left(\frac{^{88}Sr}{^{86}Sr}\right)_{sample}$ are calculated after correction for blanks and Kr- and Rb-isobaric interferences, $\left(\frac{^{92}Zr}{^{90}Zr}\right)_{measured}$ is the background-corrected ratio, and $\left(\frac{^{92}Zr}{^{90}Zr}\right)_{ref}$ is the international convention values of 0.333333 (IUPAC, De Laeter et al., 2003).

The radiogenic Sr ratio ($^{87}Sr/^{86}Sr$) is calculated by normalizing the $\left(\frac{^{87}Sr}{^{86}Sr}\right)_{corr}$ ratio to a fixed $\left(\frac{^{88}Sr}{^{86}Sr}\right)_{ref}$ ratio (8.375209, IUPAC, De Laeter et al., 2003) termed as an internal correction:

$$^{87}Sr/^{86}Sr = \left(\frac{^{87}Sr}{^{86}Sr}\right)_{corr} \times \left[\left(\frac{^{88}Sr}{^{86}Sr}\right)_{ref} \Big/ \left(\frac{^{88}Sr}{^{86}Sr}\right)_{corr}\right]^{\ln\left(\frac{m87}{m86}\right)/\ln\left(\frac{m88}{m86}\right)} \quad (5)$$

The calculated $^{87}Sr/^{86}Sr$ ratio accounts only for the variation of radiogenic $^{87}Sr$ from $^{87}Rb$ decay (Faure and Mensing, 2005).

The $\delta^{88/86}Sr$ of each bracketing standard is presented in δ-notation, as determined by the corrected $\left(\frac{^{88}Sr}{^{86}Sr}\right)_{corr}$ ratio and the certified $^{88}Sr/^{86}Sr$ ratio of 8.37861 as the normalization value. For any given sample, the corrected $\left(\frac{^{88}Sr}{^{86}Sr}\right)_{corr}$ ratio and the average $\left(\frac{^{88}Sr}{^{86}Sr}\right)_{corr}$ ratio of two adjacent bracketing NIST SRM 987 standards were used to calculate the $\delta^{88/86}Sr$ value. We should note that the SSB is only applied on $\delta^{88/86}Sr$ and not on $^{87}Sr/^{86}Sr$.

Theoretically, one could also choose to use the $\frac{^{91}Zr}{^{90}Zr}$ ratio for the Zr correction (eq. 3 and 4). We find, however, that using the more abundant isotope $^{92}Zr$ yields better long-term precision in

$\delta^{88/86}$Sr than using $^{91}$Zr, in agreement with Liu et al. (2012). Therefore, all $\delta^{88/86}$Sr values reported in this study are based on $^{92}$Zr correction. The $^{87}$Sr/$^{86}$Sr ratio is not affected by the chosen Zr isotope because of the internal correction in equation 5.

**2.3 Sr concentration**

The Sr concentration of purified samples was determined by MC-ICP-MS using the calibration curve established by Sr solutions (SCP Science PlasmaCAL Sr calibration standard in 5% HNO3) measured at six different concentrations (1ng/g, 10ng/g, 25ng/g, 50 ng/g, 100ng/g, 200ng/g). Each calibrating and sample solution was measured for 20 cycles with an integration time of 4.194 s in each cycle for $^{88}$Sr. Three blind tests (20 ng/g, 75 ng/g, and 150 ng/g) show that the measurement uncertainties were within 2%.

**2.4 Tests of different matrix and contaminations**

In order to investigate the impact of different matrices and the potential for different sources of contamination to influence Sr isotope measurements, we conducted a series of tests using the NIST 987 standard doped with Zr solution. These analyzed NIST SRM 987 standards were treated as "samples" and each was normalized to the average of its two adjacent bracketing standards following the SSB method. Thus, zero represents the "true" value of $\delta^{88/86}$Sr for our tested samples. We chose NIST 987 with concentrations of 100 ng/g Sr and 200 ng/g Zr as our bracketing standard to achieve the optimal signal intensity for both Sr and Zr isotopes. We noted that previous studies used bracketing standards with different concentrations (e.g., 150 ng/g Sr and 300 ng/g Zr in Liu et al., 2012). The actual choice of Sr and Zr concentrations of bracketing standards is likely dependent on different machine configurations, but it does not affect the objective of our designed experiments. In our tests, each sample was analyzed at least three times to gain statistical meaningfulness.

*1) Sr concentration mismatch tests*: To evaluate how Sr concentration mismatch of the sample and bracketing solutions affects Sr isotope results, we analyzed seven different concentrations (12.5 ng/g, 25 ng/g, 37.5 ng/g, 50 ng/g, 75 ng/g, 100 ng/g, 125 ng/g) of standard, while Zr concentration was held constant at 200 ng/g.

*2) Zr concentration mismatch tests*: To assess whether and how the Zr concentration mismatch of sample and bracketing solutions could affect Sr isotope values, we tested standards doped with Zr solution for five different Zr concentrations (50 ng/g, 100 ng/g, 150 ng/g, 200 ng/g, and 300 ng/g) and with Sr concentrations held constant at 100 ng/g.

*3) Acid molarity mismatch tests*: To test whether variability in acid strength could affect the sensitivity and stability of Sr isotope measurements, five different acid molarities (0.1 N, 0.5 N, 1 N, 2 N, 3 N) were tested. The bracketing standard was in 5% $HNO_3$, which is approximately 1.12 N $HNO_3$.

*4) Cation contamination*: Ba exists in most geological or biological samples. During chromatographic elution, the Eichrom Sr-spec resin exhibits a higher degree of Ba retention, relative to other cations. To address the influence of $Ba^{2+}$ on Sr isotope analyses, $Ba^{2+}$ (SCP Science PlasmaCAL Ba calibration standard in 5% $HNO_3$) was added to the NIST SRM 987 standard. The final concentrations of $Ba^{2+}$ in the analyzed solutions are 50 ng/g, 100 ng/g, and 200 ng/g.

Although the Sr-spec resin is efficient in separating $Ca^{2+}$, Ca background in some lab settings may be high and therefore we also evaluated the potential effects of $Ca^{2+}$ contamination on Sr isotope analyses. We therefore added $Ca^{2+}$ (SCP Science PlasmaCAL Ca calibration standard in 5% $HNO_3$) to the bracketing standard to achieve final $Ca^{2+}$ concentrations of 12.5 ng/g, 25 ng/g,

50 ng/g, and 100 ng/g. The Sr and Zr concentrations of all sample solutions for $Ba^{2+}$ and $Ca^{2+}$ tests were kept the same as the bracketing standard.

Resin uptake of Sr can be 'poisoned' by $K^+$ that significantly reduces $Sr^{2+}$ uptake efficiency (Dietz et al., 1992). NIST 987 bracketing standard was added with three different levels of $K^+$ (50 µg/g, 100 µg/g, and 200 µg/g; SCP Science PlasmaCAL K calibration standard in 5% $HNO_3$) and passed through Sr columns. The purified Sr solutions were dried, re-dissolved in 5% $HNO_3$, and their Sr isotope composition was analyzed.

*5) Anion contamination:* To evaluate matrix interference caused by anions, two anions, $SO_4^{2-}$ and $Cl^-$, were tested. $Na_2SO_4$ and $NaCl$ (TraceMetal Grade, Fischer Scientific) were added into NIST 987 solution to achieve 50 ng/g, 100 ng/g, and 200 ng/g in $SO_4^{2-}$ and $Cl^-$ concentrations, respectively. The mixtures were dried, redissolved in 8 $HNO_3$, and then processed through Sr columns to separate $Na^+$. The purified mixtures were then doped by Zr, and the final concentrations were 100 ng/g for Sr and 200 ng/g for Zr.

## 3. RESULTS AND DISCUSSIONS

### 3.1 Sr signal intensity, instrumental drift, and the Sr isotope values of reference materials

The bracketing standard ([Sr] = 100 ng/g and [Zr] = 200 ng/g) yielded ~33V signal in $^{88}Sr$, marking a ~10% increase in $^{88}Sr$ signal compared to a Zr-free solution with identical Sr concentration under the same instrumental setup. This signal enhancement has been previously documented in several other element-doped isotope systems (e.g., Barling and Weis, 2008; van den Boorn et al., 2009) and could be explained by analyte excitation mechanisms (cf. Gregoire et al., 1986). Given that changes of electron temperature and electron number density are inversely correlated to the ionization potential of the interferent (Hanselman et al., 1994), the introduction of Zr tends to cool the plasma and centralize the electron regions that increase single-charged ions and electron number density in the central channel, leading to enhanced signals.

Significant instrumental drift has also been observed during the six-month period of this study (Figure 1). Figure 1-A shows the determined $^{87}Sr/^{86}Sr$ ratio based on eq. 5 and Figure 1-B shows the non-SSB corrected $\delta^{88/86}Sr$ values, where $^{88}Sr/^{86}Sr$ ratios were normalized to a certified value of of NIST SRM 987 ($^{88}Sr/^{86}Sr = 8.37861$; National Institute of Standards and Technology). These temporal drifts are thought to be caused by the degradation of Faraday cup efficiency (FCE) over instrument usage, also known as Faraday cup deterioration (FCD). Many factors could contribute to FCD, including the intensity and momentum of incoming ion beans, the material and surface geometry of the cup, and contaminant deposition on cup surface (Wieser and Schwieters, 2005; Carlson et al., 2014; Garçon et al., 2018). Such changes in FCEs have been widely reported in the TIMS literature (e.g., Holmden and Bélanger, 2010; Carlson et al., 2014; Di et al., 2021) but rarely noted in studies employing MC-ICP-MS, although it should be a general issue for all mass spectrometers using Faraday cups. In principle, the conventional SSB method is expected to effectively cancel these FCEs if one assumes they are constant between analyses. However, this assumption requires a close Sr and Zr concentration match and relatively pure matrix for high-precision $\delta^{88/86}Sr$ determination (see sections 3.3 and 3.4 below). Moreover, the FCD is random and asymmetrical, and, in reality, complete elimination of these random errors is hard to achieve by SSB. Based on our observation from one continuous session (Figure 1-A and B), the drift changes significantly after every ~80 measurements. Since the determination of $^{87}Sr/^{86}Sr$ does not apply SSB, this drift could be particularly detrimental for $^{87}Sr/^{86}Sr$ ratio. Despite SSB effectively corrected the drift for our $\delta^{88/86}Sr$ values (Figure 1-D), it is hard to tell when it fails, especially when dealing with real samples. Thus, as a precautious measure, we recommend that laboratories generating Sr isotope data not exceed 40 samples per analytical session (taking account of the need for bracketing standards) in order to ensure effective correction of this drift. We also recommend frequent analysis of other reference materials to monitor the accuracy of drift correction as well as instrument stability.

For $^{87}Sr/^{86}Sr$ ratios, the instrumental drift is determined by averaging all measured $^{87}Sr/^{86}Sr$ ratios of NIST 987 standard and normalizing to the reported TIMS value of 0.710250 (Wang et al., 2019). This offset was then applied to $^{87}Sr/^{86}Sr$ of all samples and reference materials. Repeated measurements of NIST 987 yielded a mean $^{87}Sr/^{86}Sr$ ratio of 0.710250±0.000001 ($2\sigma_{SEM}$, n=125) and a mean $\delta^{88/86}Sr$ value of 0.000±0.003‰ ($2\sigma_{SEM}$, n=125) during the period of study. Repeated analyses of IAPSO seawater yielded 0.709175±0.000002 ($2\sigma_{SEM}$, n=35) and 0.390±0.005‰ (±$2\sigma_{SEM}$, n=35). Two additional reference materials (USGS BHVO-2 and BCR-2) were analyzed. BHVO-2 yielded $^{87}Sr/^{86}Sr$ = 0.703497±0.000005 ($2\sigma_{SEM}$, n=8) and $\delta^{88/86}Sr$ = 0.254±0.011‰ ($2\sigma_{SEM}$, n=8). BCR-2 yielded 0.705018±0.000004 ($2\sigma_{SEM}$, n=8) and $\delta^{88/86}Sr$ = 0.285±0.009‰ ($2\sigma_{SEM}$, n=8). These results agree well with previously published values (Ohno et al., 2007; Monynier et al., 2010; Ma et al., 2013; Andrews et al., 2016) and correspond to a long-term external reproducibility ($2\sigma_{SD}$) of ±0.000015 and ±0.03‰ for $^{87}Sr/^{86}Sr$ and $\delta^{88/86}Sr$, respectively.

**3.2 Sr column recovery and fractionation**

Previous studies have reported that incomplete recovery of Sr during chromatographic separation could induce mass fractionation of Sr isotopes (Ohno and Hirata, 2007; de Souza et al., 2010). Unlike the DS-TIMS method (Krabbenhoft et al., 2009; Wang et al., 2019), the correction technique in the Zr-doped SSB method could not correct for chromatographically induced fractionation. To evaluate the effect of different Sr column recovery rates on $^{87}Sr/^{86}Sr$ and $\delta^{88/86}Sr$, NIST 987 samples each containing 1000 ng Sr were processed through Bio-Rad columns packed with Sr-spec resin. The final Sr-release (Milli-Q water elution) step was divided into 12 sub-steps with 0.5 mL Milli-Q water in each 'cut' (i.e., the sample derived from each sub-step). The Sr concentration of each cut was analyzed (Figure 2) and the calculated Sr yield was greater than 99%. Due to low Sr concentrations at the beginning and end of the elution, we combined the first four and the last six cuts into two samples while leaving the fifth and sixth cuts intact for Sr isotope analyses. The four samples reflecting these different elution stages were dried, re-dissolved in 5% $HNO_3$, and then doped with Zr to match the Sr and Zr concentrations of the bracketing standard. As shown in Figure 2, heavier Sr isotopes were preferentially released during earlier stages of elution, resulting in heavier $^{87}Sr/^{86}Sr$ and particularly $\delta^{88/86}Sr$ values, consistent with the observation by Ohno and Hirata (2007). Based on this observed isotopic fractionation pattern, a > 95.5% recovery rate must be achieved in the purification procedure to guarantee a high-precision Sr isotopic determination ($2\sigma_{SD}$: 0.000015 in $^{87}Sr/^{86}Sr$ and 0.03 in $\delta^{88/86}Sr$). Previous studies recommended using 4 mL Milli-Q water (or 0.05 M $HNO_3$) for the final Sr release step (De Muynck et al., 2009; Neymark et al., 2014). With 1000 ng of Sr, 4mL Milli-Q water released 96.9% of total Sr. However, we found that for samples with 500 ng Sr, the elution of 4 mL Milli-Q water could only yield 93.4% of Sr. Thus, we suggest that the final Milli-Q water elution step should be increased to at least 6 mL to ensure more complete Sr recovery (Table 2). This recommendation is based on Eichrom Sr-spec Resin in size 100–150 μm. We note that smaller particle size resin (e.g., 20–50 μm) may generate a narrower elution band at the cost of slower gravity flow rates and may require lower volumes of Milli-Q water.

### 3.3 Sr concentration matching

Seven NIST 987 solutions with different Sr concentrations but the same Zr concentrations (200 ng/g) were analyzed alongside the bracketing standard ([Sr] = 100 ng/g and [Zr] = 200 ng/g) to evaluate how differences in Sr concentration between samples and bracketing standards affect Sr isotope ratios (Figure 3). This experiment shows that samples with lower Sr concentrations exhibit larger offset from the "expected" $\delta^{88/86}$Sr value (0 ‰), whereas $^{87}$Sr/$^{86}$Sr values do not exhibit obvious offset (0.71025). The variations in both $^{87}$Sr/$^{86}$Sr and $\delta^{88/86}$Sr values decrease with increased Sr concentration and signal intensity. We found that a Sr concentration mismatch of more than 50% leads to significant degradations in both precision and accuracy of the measurement. While the diminished precision can be expected from lower signal intensity, the observed inaccuracy can be attributed to two potential factors: 1) over-correction of the instrumental drift via SSB and 2) isobaric interference on $^{86}$Sr relative to $^{88}$Sr when there is a mismatch between concentrations of samples and standards. As discussed above, bombarding ion intensity accounts for one of the main sources of cup deterioration. Less concentrated Sr solutions tend to lead to smaller instrumental drift. The corrected drift with the bracketing standard is larger than the "actual" instrumental drift caused by samples with lower Sr concentrations, which can result in negative values for those samples.

Moreover, the effect of isobaric interference is exaggerated when there is a concentration mismatch between samples and standards for particular elements of interest. In the case of Sr, $^{86}$Kr can cause isobaric interference at mass 86 in the measured Sr solution. The effect of this interference on $\delta^{88/86}$Sr varied with different extents of Sr concentration mismatch and can be modelled as follows (Zhu et al., 2002; Liu et al., 2014):

$$\delta^{88/86}\text{Sr} = f \times \frac{(C-1)}{(C+f)} \times 1000 \quad (6)$$

where C is the ratio of Sr concentration in the sample solution over that in the standard solution, and $f$ is the fractional contribution of the interfering ion at mass 86 for a fixed concentration of Sr in the standard solution. The fundamental assumption involved in this model is that the sample and standard solutions should have identical "true" $\delta^{88/86}$Sr values, which is valid in our case. As shown in Figure 4, even with small interference at mass 86, the effect on $\delta^{88/86}$Sr can be significant, particularly in cases in which the sample Sr concentration is much lower than that of the standard. Based on our measured data, the best fit of *f* is 0.00004, which is equivalent to ~0.0014 V on mass 86. This suggests that even a small interference on $^{86}$Sr (~0.0014 V) may lead to large $\delta^{88/86}$Sr offsets in cases in which samples and standards have disparate Sr concentrations. Therefore, for any accurate $\delta^{88/86}$Sr determination, Sr concentrations in sample and standard solutions should be closely matched. Moreover, these results indicate that correction for $^{86}$Kr is also indispensable.

### 3.4 Zr concentration matching

To assess the effect of different Zr concentrations on measured Sr isotope ratios, the same NIST 987 solution was doped with various amounts of Zr, yielding five different Zr concentrations which were measured in conjunction with bracketing standards (Figure 5). For samples with lower Zr concentrations (< 200 ng/g), a negative offset of ~0.1 in corrected $\delta^{88/86}$Sr values is observed. For samples with higher Zr concentration (>200 ng/g), little or no offset is observed. No samples exhibited obvious offsets in their $^{87}$Sr/$^{86}$Sr ratios. With the exception of samples with the lowest Zr concentrations (50 ng/g), the variations in $^{87}$Sr/$^{86}$Sr and $\delta^{88/86}$Sr for these samples are low and consistent, which suggests that instrumental isotope discriminations are similar in Sr and Zr and can be corrected properly with the exponential law (eq. 3 and 4; Liu et al., 2012). Since there is little isobaric interference on mass 90 and 92, we attribute the observed negative offsets in samples with low Zr concentration to overcorrection introduced by SSB.

### 3.5 Acid molarity matching

Purified Sr is normally dissolved in diluted acid (e.g., 5% HNO$_3$) before introduction to the plasma. But certain factors, such as evaporation during the course of an analytical session, can potentially lead to disparity between the acid molarity of samples and standards (Teng and Yang, 2014). This acid mismatch-induced artifact has been previously reported in different metal isotope systems (e.g., Malinovsky et al., 2003; Liu et al., 2014; Teng and Yang, 2014; Oelze et al., 2016). We evaluate how such mismatch affects $^{87}Sr/^{86}Sr$ and $\delta^{88/86}Sr$ by bracketing NIST 987 standards in five different acid molarities, relative to the same standard in 5% $HNO_3$ (~1.12N). Although samples with weaker acid (e.g., ≤1N) exhibit no large shift in accuracy within the measurable limit, a negative correlation was observed between the corrected $\delta^{88/86}Sr$ value and acid molarity in that samples in strong acid yielded lower precision and were associated with a larger shift toward negative values (Figure 6). In $^{87}Sr/^{86}Sr$ space, no systematic drift is observed. Our results suggest acid molarity mismatch can affect the accuracy and precision of Sr isotope analyses and should therefore be closely monitored during each analytical session, particularly those of relatively long duration.

### 3.6 The effects of concomitant cations

Sr is a widely distributed trace element. $Sr^{2+}$ can replace $Ca^{2+}$ in Ca-bearing minerals, such as carbonate, apatite, and plagioclase, because of their similar ionic radiuses. $Sr^{2+}$ can also be captured by K-feldspar to substitute for $K^+$ ions (Faure and Mensing, 2005). Its wide distribution leads to a broad range of matrix elements when processing natural samples and the purification of Sr largely depends on the separation ability of chromatographical resin. Over the past decades, the conventional cation-exchange resin (i.e., Bio-Rad AG50) has been widely used for Sr purification (e.g., Faure and Powell, 1972). Recent years have seen the popularity of crown ether-based Sr-spec resin (4,4'(5')-di-t-butylcyclohexano 18-crown-6; Horwitz et al., 1992; Vajda et al., 1992). Ohno and Hirata (2007) compared these two resins and concluded that the Sr-spec resin could provide the higher Sr recovery rate that is crucial for $\delta^{88/86}Sr$ measurements. Despite its good performance in cation separation, Sr-spec resin shows a relatively high retention of barium (Horwitz et al., 1992). Scher et al. (2014) investigated the $Ba^{2+}$-purifying capability of Sr-spec resin and found that the 125 μL Sr-spec resin bed could purify sample solutions with $Ba^{2+}$ concentrations of up to 1 μg/g without significantly impacting measured Sr isotope values. Here, we further investigated how the presence of $Ba^{2+}$ influences Sr isotopes by adding $Ba^{2+}$ to solutions containing the NIST 987 standard. Since most geological samples examined for Sr isotopes are carbonate, we also added $Ca^{2+}$ to the NIST 987 standard to simulate potential Ca contamination during sample preparation.

As shown in Figure 7-A, the presence of $Ba^{2+}$ in Sr solutions significantly offsets $^{87}Sr/^{86}Sr$ and $\delta^{88/86}Sr$ toward negative values and the error becomes progressively larger with the increase of $Ba^{2+}$ concentration. Although the Zr correction improves the $\delta^{88/86}Sr$ precision, it does not improve the $^{87}Sr/^{86}Sr$ measurement, and it only partially compensates the offsets of both values. Although it was first thought that the presence of $Ca^{2+}$ does not influence Sr isotopic analyses (Yang et al., 2011), our results show that both $^{87}Sr/^{86}Sr$ and $\delta^{88/86}Sr$ shift systematically toward negative values (Figure 7-B). Unlike $Ba^{2+}$, this instrumental mass bias appears to be consistent and is not affected by changes in $Ca^{2+}$ concentration within the range explored for this study. Compared to the SSB-only method, the Zr correction successfully amends the $^{87}Sr/^{86}Sr$ ratio and improves the precision for $\delta^{88/86}Sr$, but the $\delta^{88/86}Sr$ values are still under-corrected even after Zr correction. Besides contaminant cation-induced changes in ionization and transmission efficiency of the analytes (Beauchemin et al., 1987), this inaccuracy could also arise from polyatomic interferences, including $^{46}Ca^{40}Ar$, $^{44}Ca^{42}Ca$, and $^{43}Ca^{43}Ca$ on mass 86; $^{44}Ca^{43}Ca$ on mass 87; and $^{44}Ca^{44}Ca$ on mass 88. Our tests reveal that a matrix effect on the accuracy and precision of $^{87}Sr/^{86}Sr$ and $\delta^{88/86}Sr$ measurements exists for both $Ba^{2+}$ and $Ca^{2+}$. Thus, high-efficiency separation of Sr from other matrix elements is needed. For natural samples with low Ba content (e.g., seawater), we suggest loading samples into columns with 1 mL 8M $HNO_3$ and follow this with an 8 mL elution of 8M $HNO_3$, as our results show no detectable $Ba^{2+}$ (on mass 138) after these procedures (Table 2). For treatment of barite samples or groundwater samples with high Ba concentrations, we recommend following the method of Scher et al. (2014).

Because the presence of potassium can cause significant reduction in Sr uptake of the resin (Dietz and Horwitz, 1993), we eluted NIST 987 standards with three different $K^+$ concentrations through the columns and measured the resulting Sr isotope composition to evaluate the extent to which $K^+$ can 'poison' the Sr-spec resin and affect Sr isotope analyses. After normal elution procedures, no detectable $K^+$ (on mass 39) was recovered from the purified Sr solution. Except for the solution with 50 μg/g $K^+$, the $^{87}Sr/^{86}Sr$ and $\delta^{88/86}Sr$ values of the analyzed solutions were significantly skewed toward negative values (Figure 7-C). After SSB and Zr correction, both values still display large deviation from true values confirming fractionation may have occurred during the Sr separation procedure due to resin poisoning. Therefore, for samples with extremely high abundance of K, a pre-precipitation of Sr with calcium phosphate or calcium oxalate may be needed to reduce the K interference (Dietz and Horwitz, 1993). The Sr-spec resin performance should not be affected by the existence of K for most of water samples, organic digests, and geological samples, including typical rocks and soils.

### 3.7 Anion contamination

The Sr-specific resin can efficiently separate $Sr^{2+}$ from other positively charged ions (Horwitz et al., 1992). However, this resin does not remove anionic or neutral species. It has been recently reported that the presence of contaminant anions could cause interreference in certain isotope systems (e.g., Pietruszka et al., 2006; van den Boorn et al., 2009). Here, we tested the interference of anions on Sr isotope measurements with two common anionic species, $SO_4^{2-}$ and $Cl^-$. Three different concentrations of $Na_2SO_4$ and $NaCl$ solutions were added to NIST 987 standard solutions and processed through Sr-spec columns (Figure 8). For non-Zr corrected values, $SO_4^{2-}$ caused positive offsets of the measured $^{87}Sr/^{86}Sr$ and $\delta^{88/86}Sr$ values, whereas $Cl^-$ produced negative offsets from the expected values. Deviations caused by either species are augmented with increasing anion concentrations. However, the SSB and Zr-corrected values do not exhibit obvious difference from expected values and fall within the range of targeted measurement uncertainty ($2\sigma_{SD}$: 0.000015 in $^{87}Sr/^{86}Sr$ and 0.03 in $\delta^{88/86}Sr$). This observation confirms that the Sr and Zr mass bias remains stable even in the presence of contaminant $SO_4^{2-}$ and $Cl^-$ and that the Zr-doping technique may help correct any error that follows exponential law. However, the mass bias may only remain stable below certain $SO_4^{2-}$ and $Cl^-$ concentrations (i.e., this may not be true of higher concentrations than those explored for this study) and may not be valid for other anions (Oelze et al., 2014).

## 4. CONCLUSIONS

Our study shows that high-quality analyses of Sr isotopic compositions can be achieved by MC-ICP-MS and precise $^{87}Sr/^{86}Sr$ and $\delta^{88/86}Sr$ values can be determined simultaneously from each measurement, using the Zr-doped SSB technique. The main source of measurement uncertainty is mass discrimination caused by sample introduction and ion extraction, which can be minimized by Zr correction. The SSB method can eliminate most of the observed shift related to FCD, based on the assumption that both standards and samples suffer are subject to instrumental drift. However, instrumental drift may be detrimental to the precise determination of $^{87}Sr/^{86}Sr$ that does not require SSB. The Sr-spec resin preferentially releases heavier Sr isotopes and therefore low Sr recovery rates (i.e. <95.5%) will lead to elevated values in $^{87}Sr/^{86}Sr$ and $\delta^{88/86}Sr$. Moreover, the Zr-doped SSB method requires a close match in concentrations of Sr and Zr as well as acid molarity between samples and bracketing standards. Contaminant cations, including $Ba^{2+}$, $Ca^{2+}$, and $K^+$, can cause large variations and deviations from real values, which cannot be corrected by Zr. The degree and direction of the offsets largely depends on the source of contamination and may vary across different instruments and laboratories (e.g., Yang et al., 2011). Correctable offsets have been found for $SO_4^{2-}$ and $Cl^-$ contaminations, although this may only be valid within the tested range and below a certain impurity level. Overall, for high-quality Sr isotope analyses, we recommend an updated chromatographic separation procedure (Table 2) to ensure complete Sr recovery and cation separation, not exceeding 40 samples per analytical session to avoid any random instrumental drift, and Sr concentration checks prior to Sr isotope measurement to ensure high rates of Sr recovery and relative homogeneity in concentrations of samples and standards.

**Acknowledgements**

We are grateful to A. Jacobson, A. Rooney, X. Gu, and Y. Di, for discussions that improved this study. This work was supported by funding from Yale University (L.G.T. and J.W.) and the Packard Foundation (N.J.P.).

**Figure captions**

**Figure 1.** Repeated measurements of NIST SRM 987 $^{87}Sr/^{86}Sr$ and $\delta^{88/86}Sr$ values over 6 months. (A) Raw $^{87}Sr/^{86}Sr$ ratios before drift-correction; (B) $\delta^{88/86}Sr$ values determined by normalizing to the certified standard $^{88}Sr/^{86}Sr$ ratio of NIST SRM 987 without SSB to show the effect of instrumental drift on $\delta^{88/86}Sr$; (C) $^{87}Sr/^{86}Sr$ ratios after drift-correction; (D) $\delta^{88/86}Sr$ values after SSB. The first 97 samples were measured from one continuous session and the rest samples were measured from different sessions. Red dashed lines represent current long-term reproducibility ($2\sigma_{SD}$: $^{87}Sr/^{86}Sr = \pm 0.000015$ and $\delta^{88/86}Sr = \pm 0.03$ ‰) at the Yale Metal Geochemistry Center.

**Figure 2.** $^{87}Sr/^{86}Sr$ and $\delta^{88/86}Sr$ variations across different elution stages of chromatographic separation using 1 mL Eichrom Sr-spec resin bed (100–150μm). Red dashed lines represent the expected "true" values. Open circles represent corrected $^{87}Sr/^{86}Sr$ and $\delta^{88/86}Sr$ values. Grey circles represent Sr weight at each elution sub-step. The total weight of Sr passed through column is 1000 ng.

**Figure 3.** $^{87}Sr/^{86}Sr$, $\delta^{88/86}Sr$, and $^{88}Sr$ intensity variations for samples with the same Zr concentration (200 ng/g) and different Sr concentrations. Box plot illustrates the distribution of $\delta^{88/86}Sr$ values. Open circles represent Zr-corrected values. Red circles represent $^{88}Sr$ intensity. Red dashed lines represent the expected "true" values. Grey areas denote current long-term reproducibilities for $^{87}Sr/^{86}Sr$ and $\delta^{88/86}Sr$.

**Figure 4.** $\delta^{88/86}Sr$ variations impacted by interference at mass 86, associated with variation in extent of Sr concentration mismatch between samples and standards. *f* denotes the fractional contribution of the interfering ion at mass 86 for a fixed concentration of Sr in the standard solution.

**Figure 5.** $^{87}Sr/^{86}Sr$, $\delta^{88/86}Sr$, and $^{90}Zr$ intensity variations for samples with the same Sr concentration (100 ng/g) and different Zr concentrations. Box plot illustrates the distribution of

$\delta^{88/86}$Sr values. Open circles represent Zr-corrected values. Red circles represent $^{90}$Zr intensity. Red dashed lines represent the expected "true" values. Grey areas denote current long-term reproducibilities for $^{87}$Sr/$^{86}$Sr and $\delta^{88/86}$Sr.

**Figure 6.** $^{87}$Sr/$^{86}$Sr and $\delta^{88/86}$Sr variations for samples with different acid molarities. Box plot illustrates the distribution of $\delta^{88/86}$Sr values. Red dashed lines represent the expected "true" values. Grey areas denote current long-term reproducibilities for $^{87}$Sr/$^{86}$Sr and $\delta^{88/86}$Sr.

**Figure 7.** $^{87}$Sr/$^{86}$Sr and $\delta^{88/86}$Sr variations for samples with different concentrations of cation contaminants, including (A) $Ba^{2+}$, (B) $Ca^{2+}$, and (C) $K^+$. Box plot illustrates the distribution of $\delta^{88/86}$Sr values. Open circles represent Zr-corrected values. Red crosses represent values without Zr correction. Red dashed lines denote the expected "true" values.

**Figure 8.** $^{87}$Sr/$^{86}$Sr and $\delta^{88/86}$Sr variations for samples with different concentrations of anion contaminants, including (A) $SO_4^{2-}$ and (B) $Cl^-$. Box plot illustrates the distribution of $\delta^{88/86}$Sr values. Open circles represent Zr-corrected values. Red crosses represent values without Zr correction. Red dashed lines denote the expected "true" values.

**Table 1 Instrument setting and operation parameters**

| Mass spectrometer setup | |
|---|---|
| MC-ICP-MS | Neptune Plus |
| RF Power | 1200 W |
| Guard electrode | On |
| Interface cones | Nickel samples cone<br>Nickel "H" skimmer cone |
| Plasma Ar gas flow rate | 16 L/min |
| Auxiliary Ar gas flow rate | 0.8 L/min |
| Ar carrier gas flow rate | 0.96 L/min |
| Stabilizing $N_2$ gas flow rate | 1.61 L/min |
| Sample uptake rate | 75 μL/min |
| Cup and amplifier configurations | L4 with $10^{12}\,\Omega$ – 83Kr<br>L3 with $10^{11}\,\Omega$ – 84Sr / 84Kr<br>L2 with $10^{12}\,\Omega$ – 85Rb<br>L1 with $10^{11}\,\Omega$ – 86Sr / 86Sr<br>C  with $10^{11}\,\Omega$ – 87Sr /87Rb<br>H1 with $10^{11}\,\Omega$ – 88Sr<br>H2 with $10^{11}\,\Omega$ – 90Zr<br>H3 with $10^{11}\,\Omega$ – 91Zr<br>H4 with $10^{11}\,\Omega$ – 92Zr |
| Desolvating device | Apex Omega |
| **Data acquisition parameters** | |
| Acquisition type | Static |
| Lens and Zoom optics | Optimized for maximum signal intensity and stability in each analytical session |
| Resolution | ~300 |
| Integration time | 4.194 s |
| Acquisition protocol | 1 block 50 cycles |

**Table 2 Chromatographic separation scheme**

| Sample with 500 ng Sr and low Ba concentration | | | |
|---|---|---|---|
| Eichrom Sr-Spec resin (100–150 μm, 1 mL resin bed) | | | |
| **Separation stage** | | **Solution matrix** | **Volume (mL)** |
| Alternating three times | Pre-clean | 8N $HNO_3$ | 1 |
| | Pre-clean | MQ water | 1 |
| Equilibrate | | 8N $HNO_3$ | 1 |
| Sample loading | | 8N $HNO_3$ | 1 |
| Elution | | 8N $HNO_3$ | 8 |
| Elution | | 3N $HNO_3$ | 4 |
| Result | | MQ water | 6 |

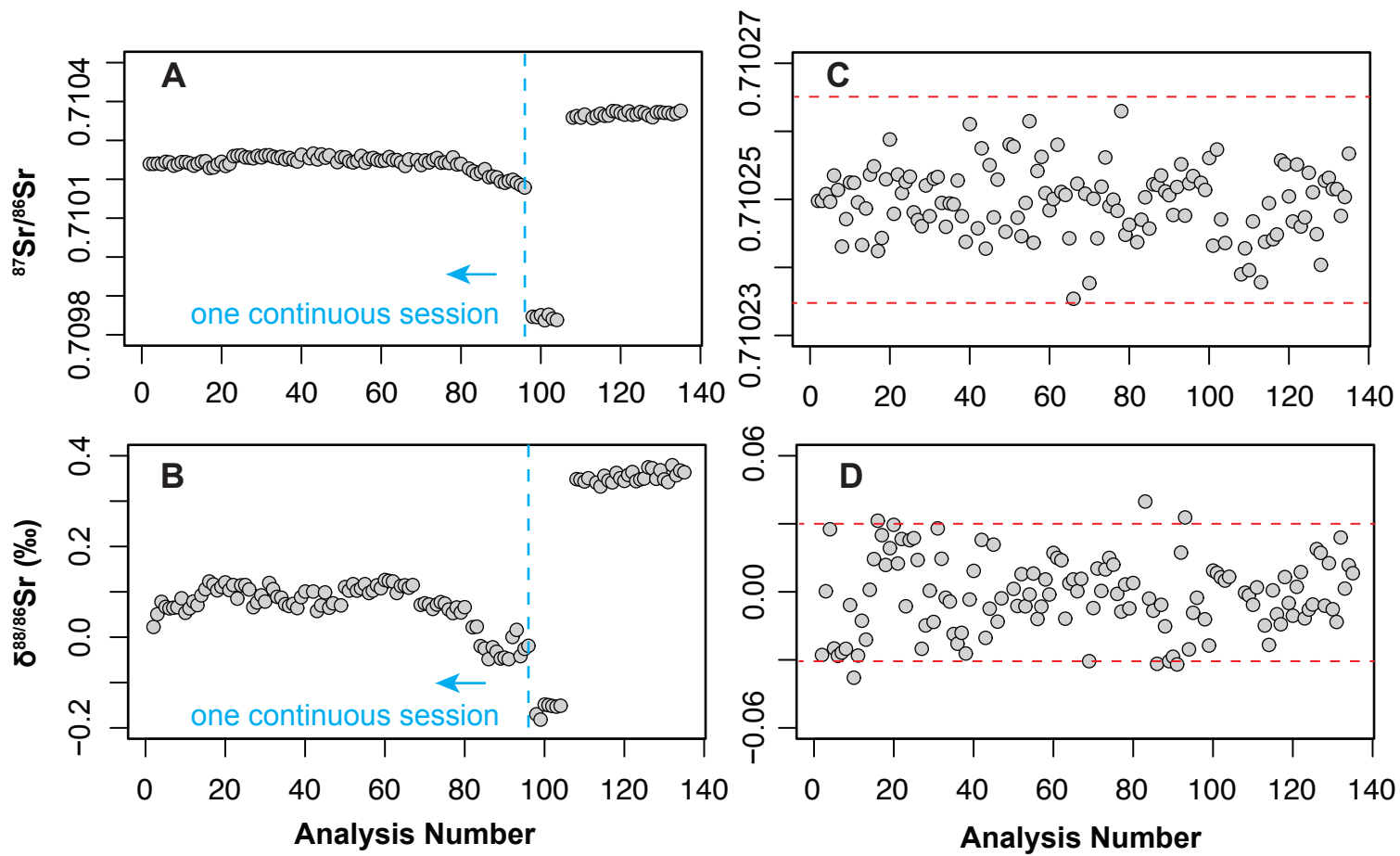

**Figure 1**

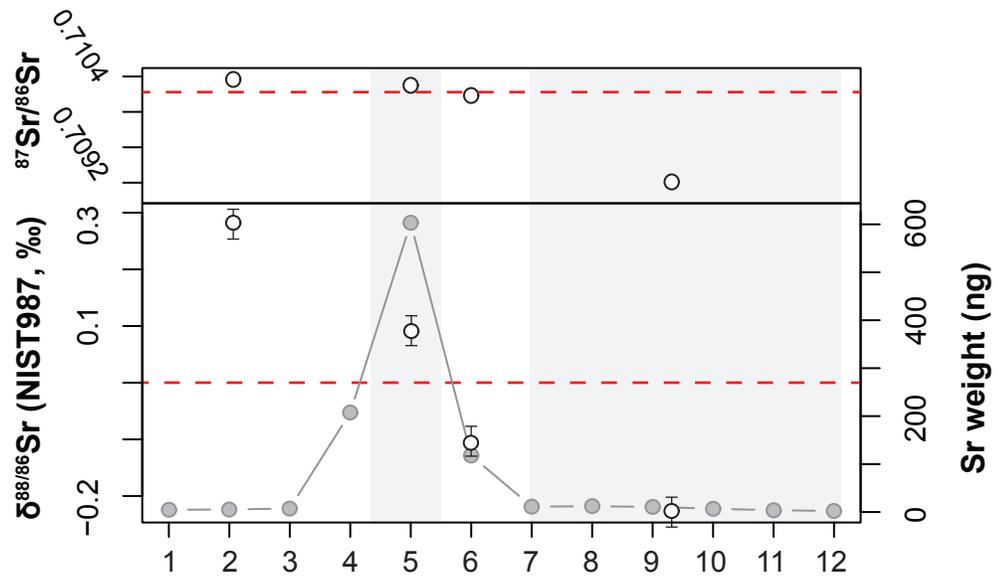

**Figure 2**

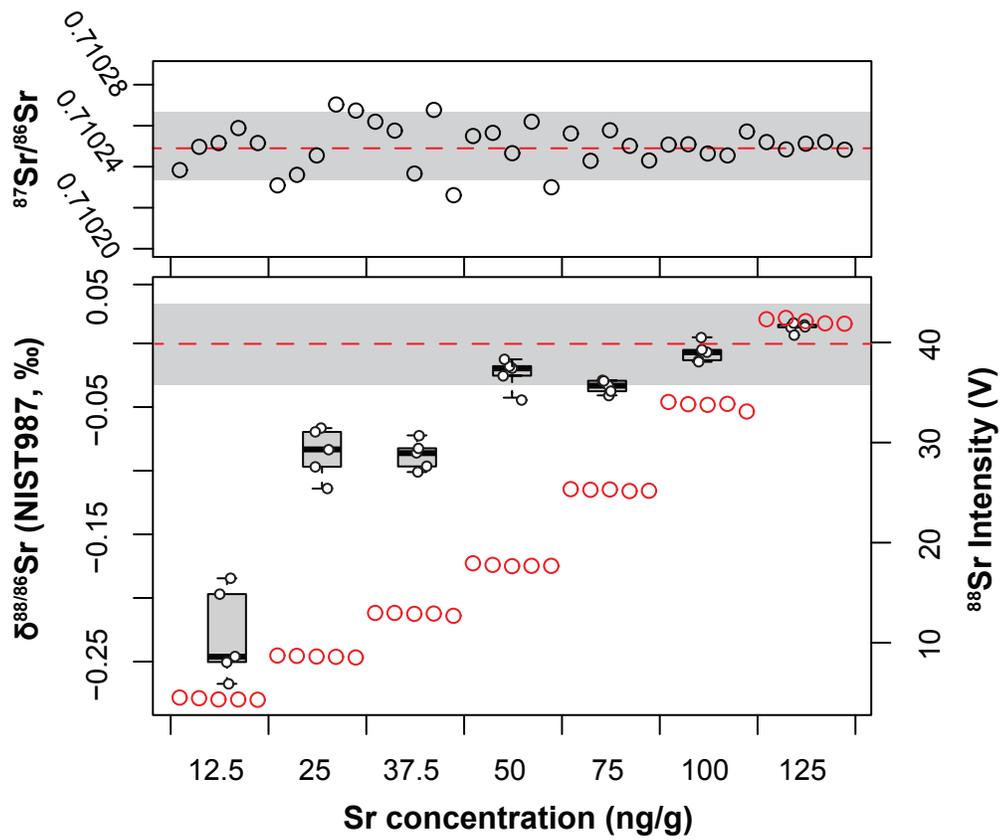

Figure 3

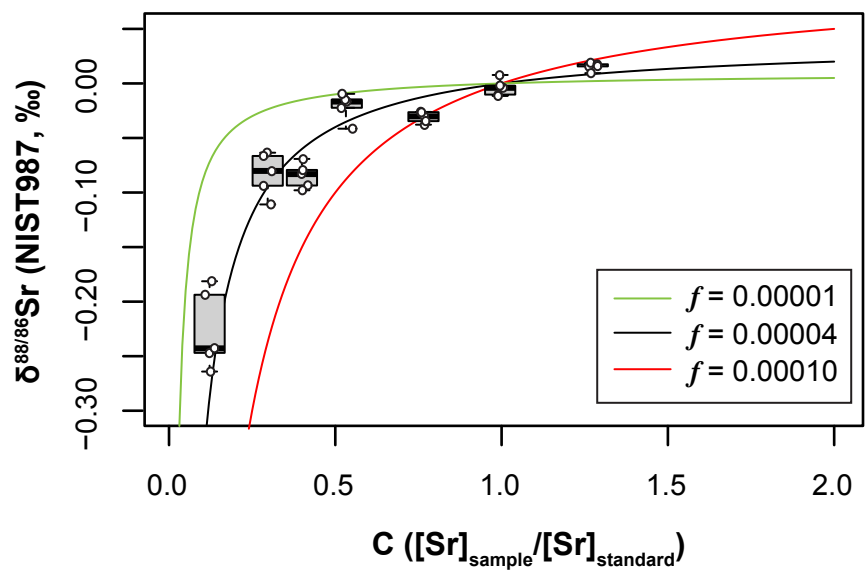

**Figure 4**

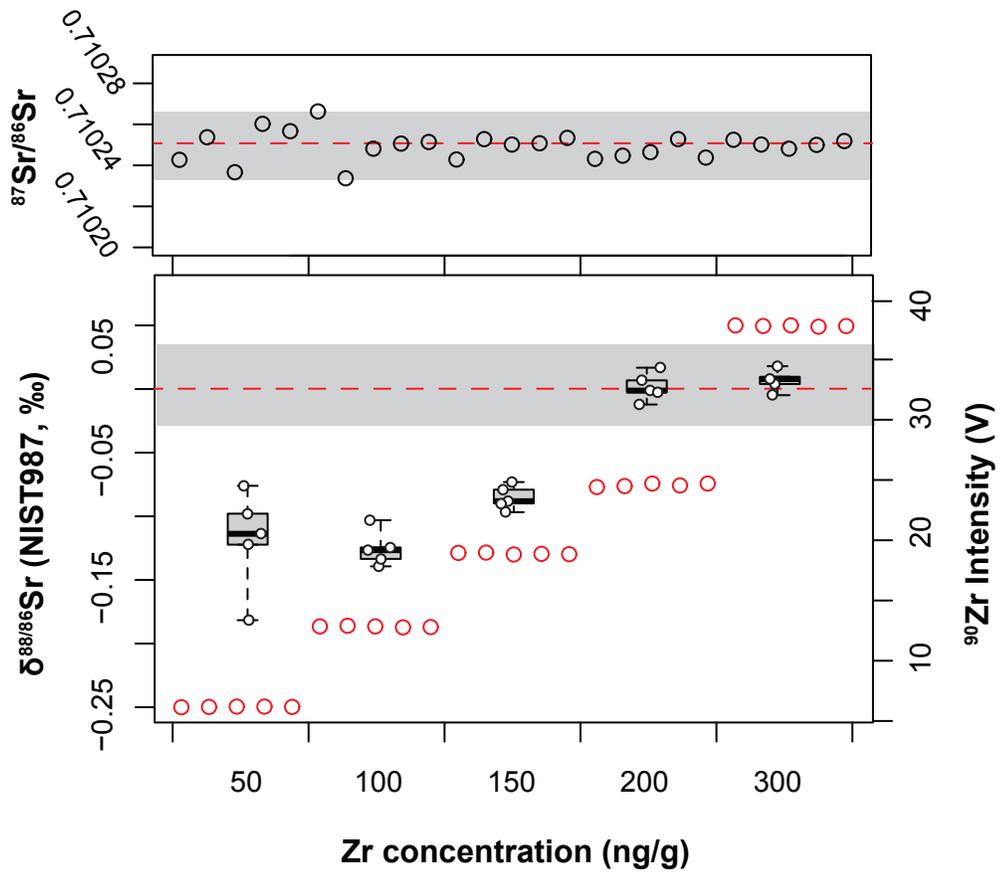

Figure 5

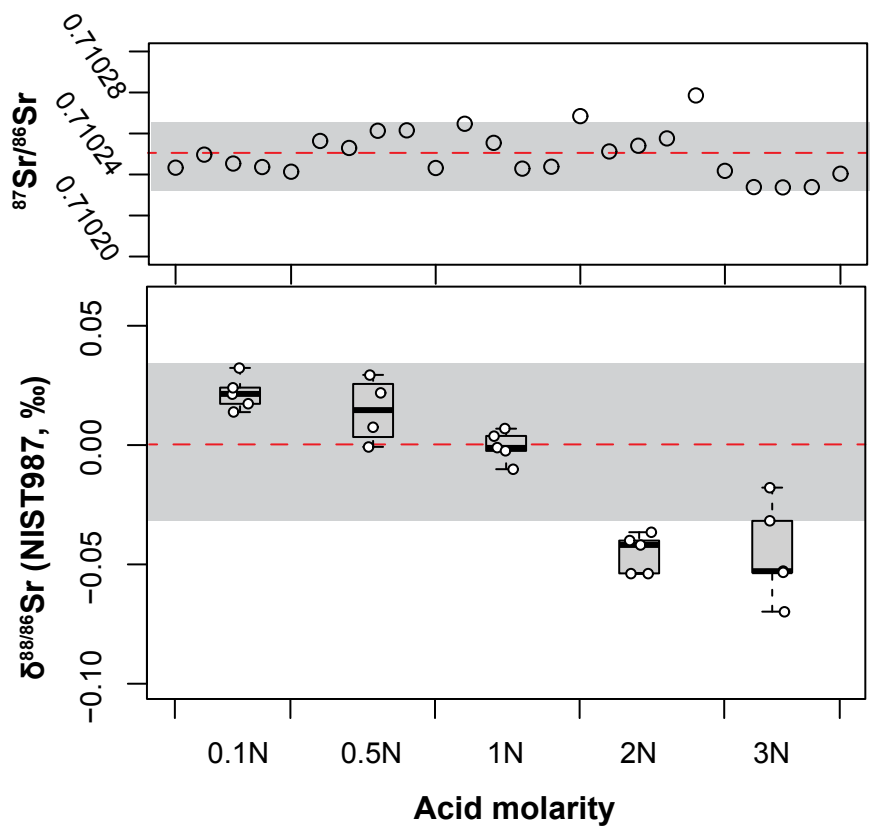

Figure 6

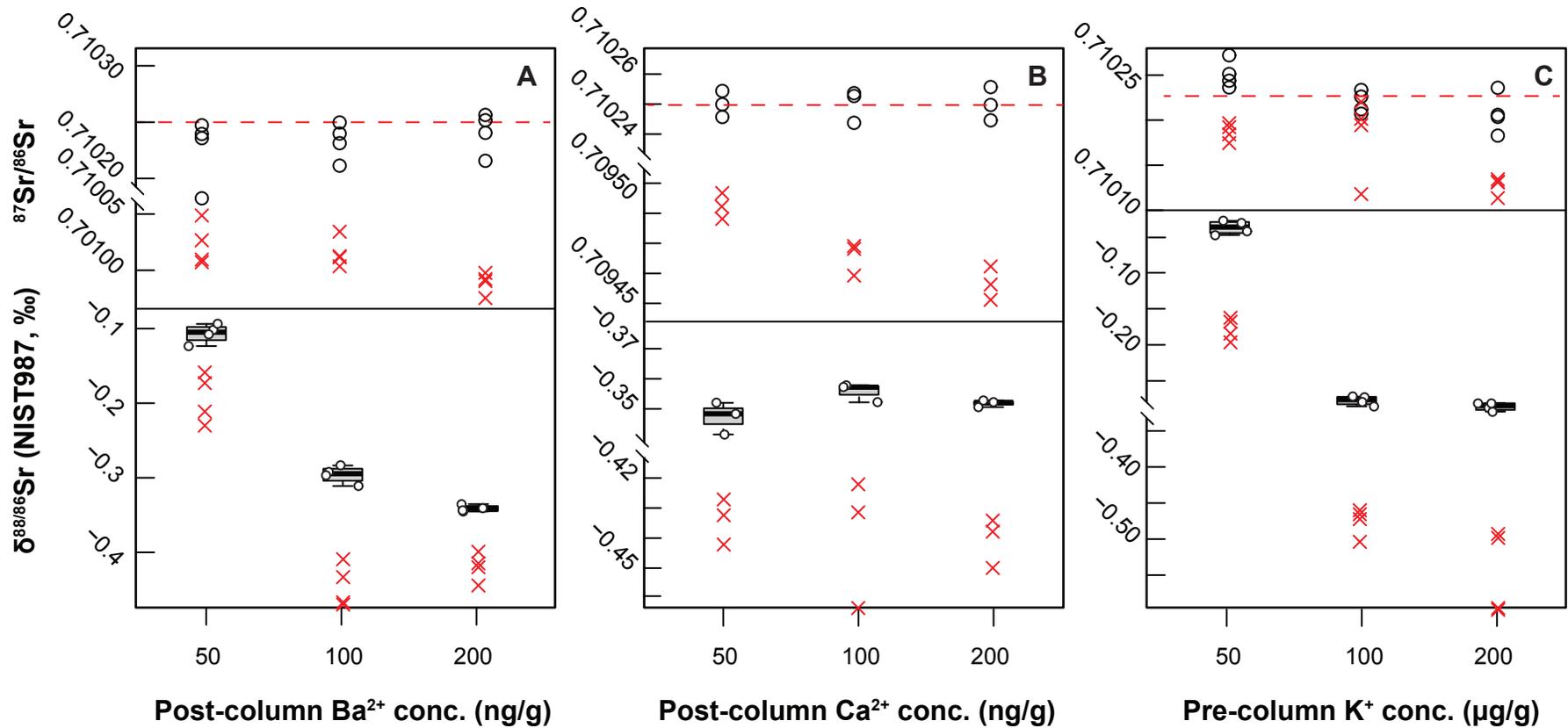

Figure 7

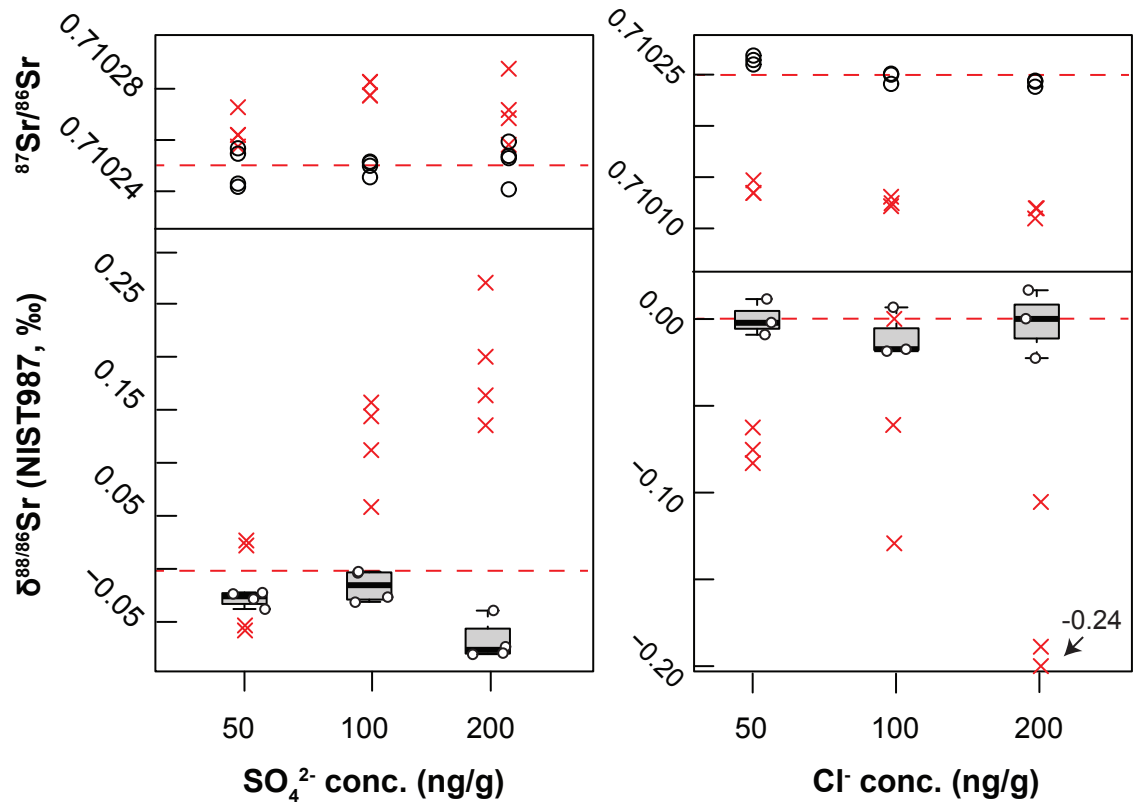

Figure 8